\documentclass[%
 reprint,
 amsmath,amssymb,
 aps,
pra,superscriptaddress
]{revtex4-2}

\usepackage{graphicx}% Include figure files
\usepackage{dcolumn}% Align table columns on decimal point
\usepackage{bm}% bold math
\usepackage{xcolor}
\usepackage{amsmath}
\usepackage[normalem]{ulem}

\usepackage{lineno}

\begin{document}

\preprint{APS/123-QED}

\title{Strong-Field Double Ionization in a Three-Electron Atom:\\ Momentum Distribution Analysis}
\author{Dmitry K. Efimov}
\email{dmitry.efimov@pwr.edu.pl}
\affiliation{Institute of Theoretical Physics, Faculty of Fundamental Problems of Technology,
Wrocław University of Science and Technology, 50-370 Wrocław, Poland}
\author{Artur Maksymov}
\affiliation{Institute of Theoretical Physics, Faculty of Physics, Astronomy and Applied Computer Science,
Jagiellonian University, Stanis\l{}awa Łojasiewicza 11, 30-348 Kraków, Poland}
\author{Jakub Zakrzewski}
\affiliation{Instytut Fizyki Teoretycznej, Wydział Fizyki, Astronomii i Informatyki Stosowanej, Uniwersytet Jagiello\'nski, ulica Stanis\l{}awa Łojasiewicza 11, 30-348 Kraków, Poland}
\affiliation{Mark Kac Center for Complex Systems Research,
Jagiellonian University, Stanis\l{}awa Łojasiewicza 11, 30-348 Kraków, Poland}
\author{Jakub S. Prauzner-Bechcicki}
\email{jakub.prauzner-bechcicki@uj.edu.pl}
\affiliation{Jagiellonian University in Kraków, Faculty of Physics, Astronomy and Applied Computer Science,
Marian Smoluchowski Institute of Physics, Łojasiewicza 11, 30-348 Krakow, Poland}

\date{\today}% It is always \today, today,

\begin{abstract}
We study strong-field double ionization in a three-electron atom by applying a simplified, reduced-dimensionality model with three active electrons. The influence of the spin-induced symmetry of the spatial part of the wavefunction  on the final two-photoectron momentum distribution is discussed. We identify partial momentum distributions originating from different sets of spins of outgoing electrons providing in this way a quantum support connection between V-structure and direct ionization typically explained classically. Changes in the momentum distribution with increasing field amplitude obtained in our simplified model are shown to be well-correlated with experimental data known from the literature. The possible relation between the observed dependencies and different  ionization  mechanisms is discussed.
\end{abstract}

\maketitle

%\tableofcontents
\section{\label{sec:intro}Introduction}

Multiple ionization of atoms and molecules is an important branch of strong-field and attosecond physics primarily due to the fact that it serves as scientific playground to study electron-electron correlations.
The best example of such a correlation is the non-sequential multiple ionization~\cite{Lhuillier82,huilier1983multiply} with its characteristic ``knee'' in the yield versus laser intensity plot~\cite{Fittinghoff92,Kondo93,Walker94} and the V-structure in the photoelectron momentum distribution (PMD) for double ionization~\cite{staudte2007binary,rudenko2007correlated}.
When two electrons are involved in a process, the well-known three-step model~\cite{Corkum93} provides the widely accepted mechanism: (i) a tunneling ionization of a single electron, (ii) its rescattering with a parent ion, and (iii) a double ionization. When more than two electrons are involved, the first two steps are similar but, as a result, in the third step one observes double, triple, and higher ionization events, all of which exhibit correlations.

The measurement of ionized particles' momenta is a challenging task, especially when triple and higher ionization is considered~\cite{Basnayake2022,Grundmann20,Larimian20,Henrichs18,Bergues12,Mikaelsson20,Zhong20}. Yet, both numerical simulations and analytical (including semi-analytical) descriptions of the process remain a difficult task too, particularly when a quantum-mechanical explanation is requested. Even for two-electron systems, the results of full-dimensional quantum simulations are scarce~\cite{parker1998intense,parker2000time,Parker06,feist2008nonsequential,Hao14}. The methods of choice in the multiple ionization case are then reduced-dimensionality quantum models~\cite{Lein00,ruiz2006ab,prauzner2007time,prauzner2008quantum,chen2010double,eckhardt2010phase,Efimov18} or classical/semi-classical calculations~\cite{Grobe94,sacha2001triple,ho2005classical,ho2005nonsequential,ho2006plane,Emmanouilidou06,Ho07,zhou10,Emmanouilidou2011,tang13,Katsoulis2018,Peters2022,Jiang2022,Emmanouilidou2023}. In the following, we shall use the reduced-dimensionality model for three active electrons introduced earlier \cite{Thiede18}.

When one constructs theoretical models for double ionization it is usually assumed that two electrons are active only, even for targets with higher number of electrons possibly involved (eg. Ne, Ar, Kr). Consequently, it is often taken for granted that these two active electrons have opposite spins. The metastable $^3S$ state of He was analysed in that contexts to expose a clear suppression of correlated escape of electrons with the same spin as a main result~\cite{Ruiz2003,Ruiz2004,Eckhardt2008}. When a larger number of electrons is considered, the spin degrees of freedom become indispensable even though the strong-field Hamiltonian does not couple to the spin part; the spin part of the wavefunction determines symmetry of the spatial part of the function. For the sake of simplicity we restrict the discussion to three-electron atoms. By a term "three-electron atoms" we understand all multi-electron atoms from which three electrons may be involved in a interaction with a laser field in an unambiguous way with respect to their spin degrees of freedom (eg. these include B, Al,... or N, P, ...). Let us consider atoms that have $s^2p^1$ electronic configuration (i.e. B, Al, ...) - the spatial part of the wavefunction will be partially symmetric with respect to exchange of electrons, because the spin part is partially antisymmetric too. On the contrary, elements with $p^3$ electronic configurations (i.e. N, P, ...) have symmetric spin part of the ground state wavefunction and, thus, the spatial part is completely antisymmetric. We have addressed the role that is played by symmetry of the spatial part of the wavefunction in the strong-field ionization of three-electron atoms in our previous studies focusing on the triple ionization yields and three-electron momenta distributions
\cite{Thiede18,Prauzner-Bechcicki21,Efimov2021-vi} and the double ionization yields only \cite{Efimov2019-hv,Efimov2020-mo}. The present study fills up a missing part, i.e. we present momenta distribution for double ionization events from the three-electron atom.

The description of double ionization of a three-electron atom is a seemingly simpler task than dealing with triple ionization, especially when it comes to presenting electron momenta distributions~\cite{Efimov2021-vi,Jiang2022}. However, the task is not that easy when the spin degrees of freedom are properly included. We have shown that in the general case, double ionization of a three-electron atom cannot be reproduced by a straightforward combination of judiciously chosen two-active electron models \cite{Efimov2019-hv}. Thus, the two-electron (2E) PMDs of the three-electron model cannot be reproduced by combinations of PMDs from the two-electron models. Let us stress that the latter conclusion has been made solely by examination of ionization potentials diagram together with the results of two-electron ionization yields simulations.

In the following, we simulate interaction of a three-electron atom with a strong-laser field while focusing on analysis of double ionization events in terms of 2E--PMDs. Since our simulation can reveal the momentum distribution for different pairs of electrons, we ask ourselves whether they can be used for uncovering some properties of a strong-field dynamics. As we show later, there is a kind of structure in PMD, namely the so-called V-structure, that can be put into direct correspondence with the particular channel for ionization event. Another finding that concerns the asymmetry property of the PMDs is discussed extensively as well.

The work is organized as follows. First, we describe the physical and numerical models used in the simulations, then results are presented and discussed. The article ends with conclusions and acknowledgments. Atomic units are used throughout the paper unless otherwise noted.

  \section{Description of the model}
Numerical treatment of tree-electron atoms in a full configuration space is a formidable task. Therefore, we resort to a restricted-space model capable of capturing the essence of strong field electron dynamics \cite{sacha2001triple,Thiede18}. In the model, each of the electrons move along one-dimensional (1D) track inclined at a constant angle $\alpha$ ($\tan \alpha = \sqrt{2/3}$) with respect to the polarization axis of the laser pulse and forming a constant angle $\pi/6$ between each pair of tracks. The chosen geometry results from the local stability analysis applied to study the adiabatic full-dimensional potential \cite{sacha2001triple,eckhardt06,arnold2013mathematical} and has been applied previously to study three-electron dynamics yielding plausible results with respect to the triple, double and single ionization yields, and the three-electron momenta distributions~\cite{Thiede18,Efimov2019-hv,Efimov2020-mo,Prauzner-Bechcicki21,Efimov2021-vi}.

The Hamiltonian reads:
\begin{equation}
\label{hamiltonian}
    H = \sum_{i=1}^3 \frac{p^2_i}{2} + V + V_{int},
\end{equation}
where $V$ and $V_{int}$ are the atomic and interaction potentials, respectively. The atomic potential in the restricted-space has a form:
\begin{equation}
    V = -\sum_{i=1}^3 \frac{3}{\sqrt{r^2_i+\epsilon^2}}+\sum_{i,j=1;i<j}^3\frac{q^2_{ee}}{\sqrt{(r_i-r_j)^2+r_ir_j+\epsilon^2}}
\end{equation}
where a smoothing factor $\epsilon=\sqrt{0.83}$ and an effective charge $q_{ee}=1$ allow for reproduction of the triple ionization potential of Neon atom ($I_p=4.63$  a.u.). In the following, we will study atoms with $s^2p^1$ electron configuration of valence electrons, that is B, Al, Ge, ... Unfortunately, in these cases a multi-photon regime requires very low laser frequencies. Thus, in order to use standard frequency $\omega=0.06$ a.u., as a model atom with $s^2p^1$ electron configuration we investigate an artificial three-electron atom with the first three ionization potentials corresponding to Neon \cite{Efimov2019-hv,Efimov2021-vi}. In this way, the presented research will remain consistent with our previous approach to the issue of triple ionization \cite{Thiede18,Efimov2019-hv,Efimov2020-mo,Prauzner-Bechcicki21,Efimov2021-vi}, allowing us to complete the missing piece, i.e. momenta distributions for double ionization events from the three-electron atom.

The interaction term in the Hamiltonian, eq.~(\ref{hamiltonian}), is described as follows:
\begin{equation}
    V_{int} = \sqrt{\frac{2}{3}} A(t)\sum_{i=1}^3 p_i,
\end{equation}
where the vector potential is given as
$$A(t)=\frac{F_0}{\omega}\sin^2\left(\frac{\pi t}{T_p}\right)\sin\left(\omega t + \phi\right)$$
for $t\in [0,T_p]$. The laser pulse parameters are: field amplitude, $F_0$; the carrier frequency, $\omega = 0.06$ a.u.; the pulse length $T_p=2\pi n_c/\omega$; the number of optical cycles, $n_c=3$; and the carrier-envelope phase, $\phi$, which is set to zero.

As we are interested in studying the interaction of three-electron atoms with a laser pulse we need a wavefunction that describes the motion of three electrons (the nucleus is assumed to be infinitely heavy, so its dynamics is neglected). The wavefunction is composed of spatial and spin parts and, as a whole, has to be antisymmetric with respect to exchange of electrons. Depending on the spin configuration, the spatial part will have different symmetry properties with respect to the exchange of electrons (see detailed discussion in Appendix~\ref{App}). For the system described with simplified Hamiltonian, eq.~(\ref{hamiltonian}), and the atom with $s^2p^1$ electron configuration the spin part is such that two electrons have the same spin, i.e. the spin part needs to be symmetric with respect to exchange of electrons in that pair. Consequently, the spatial part needs to be antisymmetric with respect to exchange of electrons in that pair. Let us emphasize that the choice of the considered electron configuration boils down to the choice of symmetry of the wave function. The geometry of the model and the Hamiltonian both remain unchanged.

Without loss of generality we may write the wavefunction for such atom in the restricted-space model as (see eq.~(\ref{wfunct_fin})):
\begin{multline}\label{wf}
    \Psi\propto \Psi_{12}(r_1,r_2,r_3)|{\rm UUD}\rangle \\
        +\Psi_{23}(r_1,r_2,r_3)|{\rm DUU}\rangle\\
        +\Psi_{13}(r_1,r_2,r_3)|{\rm UDU}\rangle.
\end{multline}
Here $U$ and $D$ stand for electron with spin-up and spin-down, respectively. The subscripts denote the pair of electrons with respect to exchange of which the wavefunction is antisymmetric.

The Hamiltonian, eq.~(\ref{hamiltonian}), does not influence the spin part during evolution, therefore for the sake of simplicity, in numerical implementation it is enough to evolve one of the three terms on the right side of the eq.~(\ref{wf}) only.  However, as explained in detail in the Appendix, special caution must be taken with the symmetry properties of the wave function. In the following, we shall evolve the wavefunction $\Psi_{12}(r_1,r_2,r_3)$, i.e. a function that is antisymmetric with respect to exchange of electrons $1$ and $2$, but with respect to exchange of electrons $1\leftrightarrow 3$ and $2\leftrightarrow 3$ it is neither symmetric nor antisymmetric. The initial wavefunction, mimicking the ground state, is found using imaginary time propagation of the Hamiltonian, eq.~(\ref{hamiltonian}), without the interaction with the field and enforcing the proper symmetry~\cite{Thiede18}.

The numerical solution of the time-dependent Schr\"odinger equation (TDSE) with the Hamiltonian, eq.~(\ref{hamiltonian}), is essentially an extension of the method used in the two-dimensional case described {\it in extenso} elsewhere~\cite{prauzner2008quantum,Efimov2021-vi}. Here, we recall this method only very briefly. The TDSE is solved on a large 3-dimensional grid with the use of a split-operator technique and the fast Fourier transform algorithm. The goal is to obtain momenta distributions of outgoing electrons, therefore the wavefunction cannot be absorbed at edges of the grid as it is typically implemented. In our approach, we divide the whole evolution space into regions, i.e., the ``bounded motion'' and ``outer'' regions. In the ``bounded motion'' region the evolution is done without further simplifications, whereas in the ``outer'' regions the Hamiltonian is simplified, i.e. the interaction of electrons with nucleus and other electrons is successively neglected resulting in regions describing a single ion, double ion and triple ion. Thanks to these simplifications, in the ``outer'' regions the wavefunction can be represented and evolved in the momentum representation - the evolution simplifies to the multiplication by an appropriate phase factor. The transfer between the ``bounded motion'' and ``outer'' regions is done by a smooth cutting and coherent adding of the wavefunction following the procedure introduced in ~\cite{Lein00}.

At the end of the simulation, the wavefunction from ``outer'' regions can be integrated over the ``bounded'' part, thus leaving the momentum distributions corresponding to single ionization, double ionization and triple ionization. More details of the algorithm for the simulation of momentum distributions of the three-electron atom can be found in the supplementary material of the work \cite{Efimov2021-vi}.

Strong-field double ionization is a quite convoluted process. For a purpose of the present work we shall divide the processes into two categories, i.e. recollision-impact ionization (RII) and time-delayed ionization (TDI). The RII is a process in which the rescattering electron ionizes the bound electron upon collision and is usually connected with the V-structure (also named a finger-like structure) in PMD~\cite{staudte2007binary}. The TDI includes sequential ionization (independent electron releases)~\cite{Geltman85}, recollision-excitation with subsequent ionization (RESI) of all kinds~\cite{Rudenko2004,Liao2017-uh} and other possible mechanisms (eg. the slingshot non-sequential ionization~\cite{Katsoulis2018}).

The ionization yields ascribed to different ionization channels are calculated in our model by an integration of probability fluxes over the border of the appropriate regions. The RII yield is calculated by integration of the direct probability flux over the border between the regions describing a neutral atom and that for the double ion, whereas TDI yield is calculated by an integration of the flux over the border between the regions describing a single ion and double ion. Detailed procedure is described elsewhere~\cite{prauzner2008quantum,Efimov18,Prauzner-Bechcicki21}.

The grid has $n=1024$ nodes in each direction, the step in the grid is ${\rm d}r=0.195$. The time-step is ${\rm d}t=0.05$ and total number of steps amounts to $6500$.

  \section{Results and discussion}
In the following we shall analyze the final wavefunction of the system. Once the laser pulse is ended after long enough time one may plausibly assume that the wavefunction can be written as:
\begin{equation}\label{final_wf}
 \Psi = \Psi_N + \Psi_{SI} + \Psi_{DI} + \Psi_{TI},
\end{equation}
where $\Psi_N$, $\Psi_{SI}$, $\Psi_{DI}$ and $\Psi_{TI}$ are wavefunctions of an neutral atom - three bound electrons, a single ion -  a single freed and two bounded electrons, a double ion - two freed electrons and a single bounded one,  and a triple ion - three freed electrons, respectively. We refer to freed electrons as we are focused on electron dynamics. Thanks to the division into the ``bounded motion'' and ``outer'' regions in our numerical approach we have indeed access to these four wavefunctions. Before we continue our analysis, let us note that it is justified to expect that the wavefunctions, eq.~(\ref{final_wf}),
 which the final wavefunction is decomposed to, are at least approximately orthogonal to each other after enough time passed since the action of the laser field, i.e. one has a set of properties: $\langle \Psi_N| \Psi_{SI}\rangle \simeq 0, \langle \Psi_N| \Psi_{DI}\rangle \simeq 0, $ etc.
Therefore, the final observable, a PMD may be written:
\begin{equation}
|\Psi|^2 \simeq |\Psi_N|^2 + |\Psi_{SI}|^2 + |\Psi_{DI}|^2 + |\Psi_{TI}|^2.
\end{equation}
The last term, $|\Psi_{TI}|^2$, is a three-electron (3E) PMD and have been already studied elsewhere~\cite{Efimov2021-vi}. Here we concentrate on a 2E--PMD, i.e. $|\Psi_{DI}|^2$. Let us stress that our model allows us to study a three-electron atom and extract 2E--PMDs, so the influence of the third electron is treated explicitly in the evolution and may be traced in the distributions. Typically, when double ionization of multi-electron atoms is considered one refers to some kind of two-active electron model, where the influence of the third electron is neglected.

The 2E--PMD is an experimental observable. In coincidence measurements, momentum of a double ion is measured together with momentum of one of the electrons (see for example~\cite{Jiang2023}). The momentum of the remaining electron is retrieved from the momentum conservation principle. Our approach allows to shed some more light on components contributing to the observed PMDs, and originating in different processes.

The final wavefunction, eq.~(\ref{final_wf}), is a three-electron wavefunction. Its part that describes double ions, that is, $\Psi_{DI}$, has, in principle, a form of eq.~(\ref{wf}). The first term, i.e. $\Psi_{12}(r_1,r_2,r_3)$, is antisymmetric with respect to the exchange of electrons 1 and 2, while it does not possess symmetries with respect to the exchange of other pairs of electrons. The other terms posses similar properties as applied to different pairs of electrons. To obtain 2E--PMD we need to write the wavefunction in momentum representation, therefore the Fourier transform is applied. The Fourier transform does not change the symmetry properties of the wave function. Next step is to take modulus squared of the wavefunction, and here the fact that spin parts are orthogonal allows us to write it in a formal form:
\begin{multline}\label{3e_pmd}
    |\Psi_{DI}(p_1,p_2,p_3))|^2\propto \\
        |\Phi^{DI}_{12}(p_1,p_2,p_3)|^2
        +|\phi^{DI}_{23}(p_1,p_2,p_3)|^2\\
        +|\phi^{DI}_{13}(p_1,p_2,p_3)|^2.
\end{multline}
This operation changes symmetry, so for example the first term of eq.~(\ref{3e_pmd}) becomes symmetric with respect to exchange of electrons 1 and 2.
The last step in pursue to 2E--PMD is to integrate eq.~(\ref{3e_pmd}) with respect to the variable describing the bound electron. Let us assume the electron 3 is still bound, the integration over $p_3$ variable leaves us with:
\begin{multline}\label{2e_pmd}
    PMD(p_1,p_2)\propto \\
        |\Phi^{DI}_{12}(p_1,p_2)|^2
        +|\phi^{DI}_{23}(p_1,p_2)|^2\\
        +|\phi^{DI}_{13}(p_1,p_2)|^2.
\end{multline}
The first term is symmetric with respect to exchange of electrons 1 and 2 but the other two terms are not necessarily. Overall, 2E--PMD has to be symmetric with respect to the exchange of electrons because the electrons are indistinguishable. Therefore, the sum of the second and third terms in eq.~(\ref{2e_pmd}) has to be symmetric. It can be expected that when the first term of eq.~(\ref{2e_pmd}) is plotted alone it should have zeros on the $p_1=p_2$ axis and be symmetric with respect to that axis, while the other two terms if plotted separately could be asymmetric with respect to $p_1=p_2$, but the axis itself does not need to be filled with zeros.

\begin{figure*}
  \includegraphics[width=0.8\linewidth]{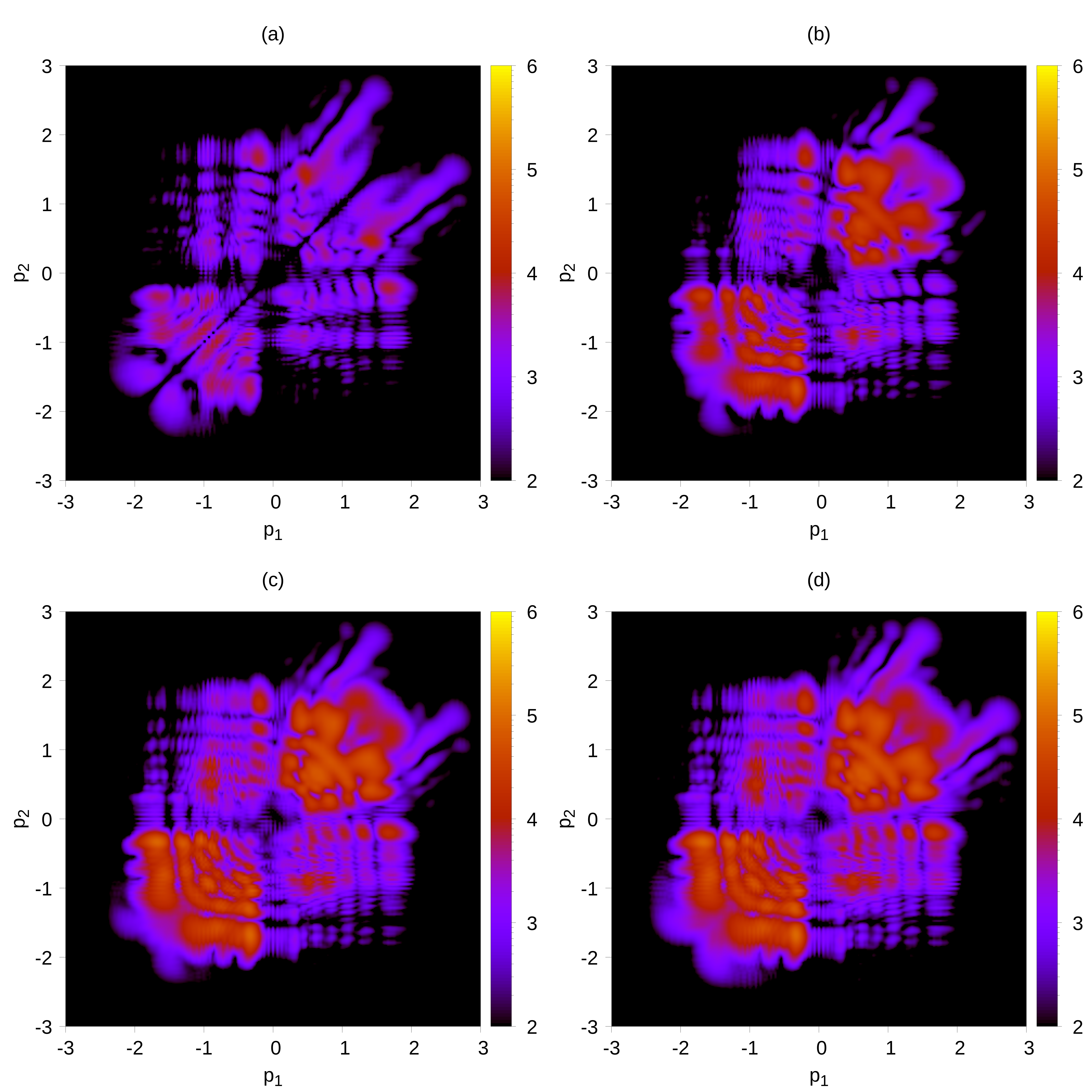}
  \caption{ Two-photoelectron momentum distributions for field amplitude F=0.12 a.u. (a) partial distribution - symmetric part, corresponding to the first term on the right side of eq.~(\ref{2e_pmd}), (b) partial distribution - asymmetric part, corresponding to the third term on the right side of eq.~(\ref{2e_pmd}), (c)  the sum of the second and third terms on the right side of eq.~(\ref{2e_pmd}), (d) the full 2E-PMD, eq.~(\ref{2e_pmd}). Atomic units are used for the scale of $p_1$ and $p_2$ axes. } 
  \label{figmom}
\end{figure*}

These symmetry expectations are confirmed in Fig.~\ref{figmom} where we present an exemplary 2E--PMD. In panel (a) the partial 2E--PMD is clearly symmetric with respect to the $p_1=p_2$ and the axis has pronounced zeros. The zeros on the axis are remnants of antisymmetry of that part of the wavefunction and have their origin in the Pauli exclusion principle (a similar effect was discussed in Helium in~\cite{Eckhardt2008}). In Fig.~\ref{figmom}(b), where the third term of the full 2E--PMD, eq.~(\ref{2e_pmd}), is depicted alone, the distribution is clearly asymmetric with respect to $p_1=p_2$ axis and has non-zero values on the axis; however, at the same time it contains elements resembling the famous V structure. The sum of the second and third terms is shown in Fig.~\ref{figmom}(c). It exhibits the V-structure undoubtedly - it is more prominent in the first quadrant because the simulations are done for a very short laser pulse (only 3 cycles long) and for a single carrier-envelope phase. The V structure is still visible in the entire 2E--PMD; see Fig.~\ref{figmom}(d). From the inspection of the full distribution, one sees that the V-structure is dominant feature, and the signal originating in the $|\Phi^{DI}_{12}(p_1,p_2)|^2$ part is barely visible. The partial PMDs cannot be measured experimentally, as the electrons are indistinguishable in the detector. The above discussion is valid exclusively within the framework of the theoretical model. However, the observed differences in the partial PMDs are the consequence of the initial-state symmetry, i.e. for atoms with totally antisymmetric spatial part (the spin part is symmetric as for $p^3$ valence shell atoms,~\cite{Efimov2020-mo}) all the partial PMDs look the same.

Furthermore, the 2E--PMD shown in Fig.~\ref{figmom}(a) can be interpreted according to the analysis presented by Staudte et al. in~\cite{staudte2007binary}. The suppression of correlated escape of electrons connected with the Pauli exclusion principle is reserved to binary collisions, whereas the observed features can be connected with recoil collisions exclusively.

The lack of symmetry in relation to the electron exchange visible in Fig.~\ref{figmom}(b), which we will call an asymmetry, follows directly from the symmetry properties of the initial state and the decomposition of the final wavefunction, see eq.~(\ref{2e_pmd}). Although the existence of such an asymmetry is clear, the degree of it does not follow directly from the definition of the model and should rather depend on the particular properties of ionization dynamics. Thus, it is an interesting task to reveal the connection between ionization and partial momentum distribution asymmetry.

The partial PMD shown in Fig.~\ref{figmom}(b) possesses a slight asymmetry, but gives no clue whether the asymmetry depends on the physical parameters such as the field amplitude. One could suspect that interaction with pulses having larger amplitudes may introduce more asymmetric partial distributions, mainly because the time-delayed double ionization, such as sequential ionization or RESI, is more pronounced for these conditions. In fact, such an increase in asymmetry was observed in our calculations.

\begin{figure}
\includegraphics[width=1.0\linewidth]{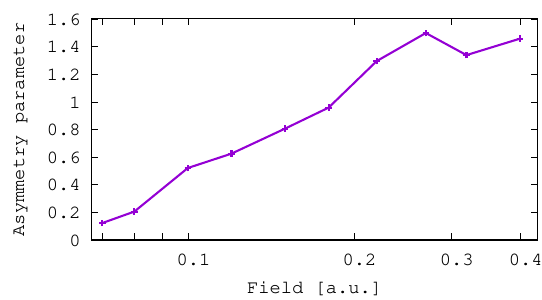}
\caption{Asymmetry parameter computed for the partial 2E--PMD (the third term of eq.~(\ref{2e_pmd})) for different field values.}
\label{figc}
\end{figure}

\begin{figure}
\includegraphics[width=1.0\linewidth]{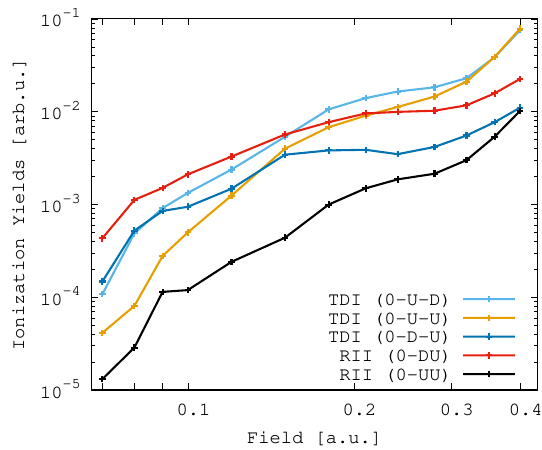}
\caption{Yields \textcolor{red}{(arb. units)} for different channels of double ionization. The TDI stands for Time-dependent ionization, that includes sequential ionization and recollision-excitation with subsequent ionization channel (RESI); the RII, or Recollisionally Induced Ionization stands for a set of channels that assume direct ionization after recollision. The order in which electrons ionize with respect to the spin degree of freedom is denoted by capital letters from left to right, ex. (0-U-D) stands for TDI in which the electron with spin up (U) is ionized first; if a direct ionization happened, the letters denoting spins are written close to each other, ex. (0-DU). }
\label{figIY}
\end{figure}

Let us introduce the asymmetry parameter of the partial 2E--PMD obtained by providing a comparison between the above-diagonal and below-diagonal regions of the distribution. In particular, we collect information about the difference of points located symmetrically in relation to the main diagonal. The asymmetry parameter $\beta$ is defined as follows:
\begin{equation}
  \beta = \cfrac{\int\limits_{-\infty}^{\infty} \int\limits_{-\infty}^{\infty} \left| |\Psi (p_1,p_2)|^2 - |\Psi (p_2,p_1)|^2\right| \,dp_1\,dp_2}{\int\limits_{-\infty}^{\infty} \int\limits_{-\infty}^{\infty} |\Psi (p_1,p_2)|^2 \,dp_1\,dp_2}.
\end{equation}
The $\beta$ parameter calculated for different field intensities for the part corresponding to the first term in eq.~(\ref{2e_pmd}) was unsurprisingly equal to 0. The dependence of $\beta$ on the field value for the third term in eq.~(\ref{2e_pmd}) is shown in Fig. \ref{figc}. Within the standard laser field amplitude interval, in which one observes the famous ``knee'' structure in double ionization yield, the $\beta$ parameter increases considerably, therefore, let us contrast it with ionization yields for different channels as a function of field amplitude, see Fig.~\ref{figIY}.

 As discussed in the previous section, the Hamiltonian, (eq.~(\ref{hamiltonian})), does not influence the spin part during the evolution, and therefore in our numerical implementation we evolve only the first term of the three terms on the right side of eq.~(\ref{wf}). Consequently, the spin degree of freedom is ``encoded'' in electron's index and it is possible to trace the yield of spin-resolved ionization channels~\cite{Thiede18}. In Fig.~\ref{figIY} we show the dependence of ionization yield on the field amplitude split into TDI and RII channels, each of which is further assigned with respect to the order of the escaping electrons. Eventually, we calculate five partial yields, i.e. RII (0-DU) -- the direct double ionization of electrons with opposite spins, RII (0-UU) -- the direct double ionization of electrons with the same spin, TDI (0-U-D) -- the time-delayed double ionization in which the first freed electron has spin up and the second has spin down, TDI (0-D-U) -- the time-delayed process with reversed sequence of escaping electrons, and finally TDI (0-U-U) -- the time-delayed ionization of electrons with the same spin. Within the discussed model, these partial yields can be easily connected with the partial 2E--PMDs depicted in Fig.~\ref{figmom} (a)--(c). Indeed, the fluxes that allow calculation of yields RII (0-UU) and TDI (0-U-U) are also responsible for filling the region of the integration grid from which the partial 2E--PMD with symmetric distribution is retrieved (see Fig.~\ref{figmom}(a)). Similarly, the fluxes allowing calculation of RII (0-DU), TDI (0-D-U) and TDI (0-U-D) yields are connected with partial PMDs depicted in Fig.~\ref{figmom}(b) and (c).

From comparison between Fig.~\ref{figc} and Fig.~\ref{figIY} it is hard to discriminate which of the identified ionization channels is responsible for changes in $\beta$. All channels that contribute to partial 2E--PMD for which the asymmetry parameter is calculated increase monotonically with increasing field amplitude. The only important fact in the ionization yield plot is the change of order, i.e. for low amplitudes the RII (0-DU) channel dominate over the two TDI channels, whereas for higher amplitudes, when sequential double ionization start to dominate the TDI (0-U-D) channel prevails.

\begin{figure*}
  \includegraphics[width=1.0\linewidth]{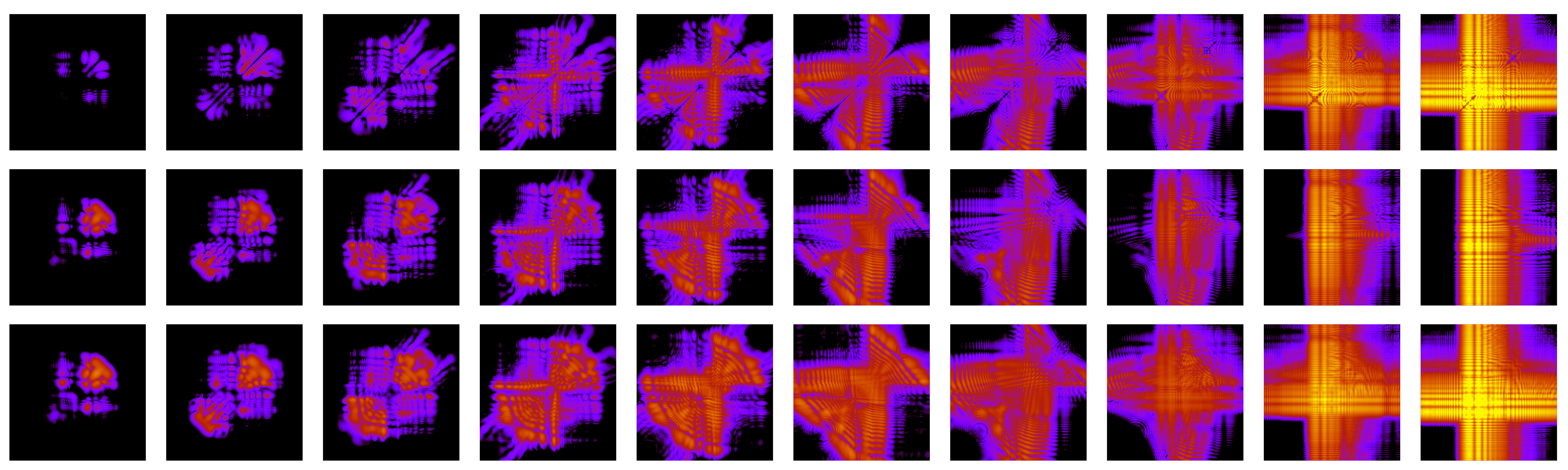}
  \caption{ Two-photoelectron momentum distributions for field amplitudes (from left to right) F=0.08, 0.10.0.12, 0.15, 0.18, 0.22, 0.27, 0.32, 0.40, 0.50 a.u. The upper, middle and lower raws correspond to (i) partial distribution - symmetric part, corresponding to the first term on the right side of eq.~(\ref{2e_pmd}), (ii) partial distribution - asymmetric part, corresponding to the third term on the right side of eq.~(\ref{2e_pmd}), (iii)  the sum of the second and third terms on the right side of eq.~(\ref{2e_pmd}). The range of each numerical box is (-3,3) a.u.}
  \label{fig_series}
\end{figure*}

It is a known fact that for amplitudes below the knee structure in the ionization yield plot the non-sequential ionization is the main channel for double ionization with its acknowledged trace in a form of the V-structure that, with increasing field intensity, changes into cross-shaped feature ~\cite{Kubel2016}. This fact is both reflected in Fig.~\ref{figc} and in the 2E--PMDs in Fig.~\ref{fig_series}. The increase of $\beta$ with increasing field amplitude may be interpreted as a signature of unequal energy sharing upon recollision.

 The connection between the probability fluxes and partial PMDs allows us to conclude that the V-structure observed in double ionization of three-electron atoms corresponds to the ionization via RII (0-UD) channel that has a distinctively higher yield than RII (0-UU) yield. It is of particular interest to recall that the V-structure in two-electron momentum distribution is known for about two decades \cite{staudte2007binary} and, indeed, is clearly associated to the direct impact channels of double ionization. However, such correspondence has been set up after simulations employing a purely classical picture of ionization-recollision processes. The observation we made nicely compliments the classical argument by adding a quantum evidence of the V-structure connection to the RII channels even in ionization of three-electron atoms.

We will place the above observation in a broader context by discussing the different types of models used so far to simulate V-shaped structures. They are necessarily limited to the two-electron case, since the modeling of multi-electron atoms is still in its infancy. The correspondence between the V-structure in 2E--PMDs to the RII (0-UD) channel is not guaranteed in the general case and its existence is sensitive to the model employed. 
Starting with the \textit{ab initio} quantum simulations of 2E-PMD,
including \cite{staudte2007binary}, all these approaches studied the ground state of Helium, which is a singlet state and, thus, used the symmetric spatial wavefunction without referring to the spin state explicitly. 
Therefore, in view of the conferred results, one should not be surprised by the presence of the V-structure in the corresponding simulated distributions. The influence of the symmetry of wavefunction on the correlated double ionization has been addressed in theses models by introducing the metastable $^3S$ state of He as an initial configuration~\cite{Ruiz2003,Ruiz2004,Eckhardt2008}. Obtained results were interpreted as a suppression of correlated escape of electrons.
Next, the classical trajectory simulations, like \cite{PhysRevLett.101.233003}, essentially use the classical phase space with symmetric initial distributions of trajectories; 
that, again, implicitly corresponds to opposite spins case. The QRS (the quantitative rescattering) approach, like in \cite{Chen20}, in essence considers coupled single-electron problems: single-particle distributions and observables obtained after solving one-electron problem are used afterwards as an external parameters for the second-electron problem. There seems to be no way to connect the single-electron wave-functions to a spatially anti-symmetric wavefunction, thus the method provides spatially-symmetric, or opposite-spins picture.

Let us now point out to one more ingredient that has its role in the formation of final PMDs, namely, RESI process. Unfortunately, within the model that we use we cannot separate RESI channel from the other possible ionisation paths. Note, we use integration of probability fluxes over borders between regions defined as atom, single, double and triple ion to calculate ionization yields and therefore RII yield can be separated from all TDI events, that is sequential ionisation and all flavors of RESI. Typically, when discussing RESI one assumes that the ionisation involves excitation of a bound electron to a singly-excited state (SES) of an ion. However, Liao et al.~\cite{Liao2017-uh} have shown in their study that there is a group of channels that involves population of a doubly excited state (DES) of an atom. Indeed, when the energy of the returning electron is not sufficient to support direct ionisation that electron can be trapped by the parent ion and share part of its energy with the bound electron to form DES. The lifetime of such states relative to autoionization is typically of order of several femtoseconds thus they may play an important role in strong-field dynamics. What is more, the relative participation of DES in ionization is noticeable for low field amplitudes and decreases monotonically with its increase, whereas the role of SES in RESI increases with increasing field amplitude~\cite{Liao2017-uh}. Although the symmetry of the excited state plays the key role in the final shape of PMD~\cite{maxwell2016,Chen17}
it is impossible, at the present stage of the analysis performed, undoubtedly to indicate its role in the change of the $\beta$ parameter. Further research in that direction is anticipated.

\section{Conclusions}
We have studied strong-field double ionization in a three-electron atom by applying a simplified, reduced-dimensionality model with three active electrons. Special attention has been paid to the spin-induced symmetry of the spatial part of the wavefunction and its influence on the final two-photoelectron momentum distribution. Partial momentum distributions originating from different sets of spins of outgoing electrons were identified. Thereafter, the distribution exhibiting the famous V-structure was undoubtedly linked with the escape of electrons with opposite spins; we have shown that it corresponds to the RII channel thus providing a quantum support for the classically grounded connection between the V-structure and direct ionization. Furthermore, the latter distribution was contrasted with the one resulting from the escape of electrons with same spins and displaying suppressed correlation. Changes in the photoelectron momentum distribution with increasing field amplitude was discussed and found coherent with data known from literature. Finally, possible relationship between the observed changes and various RESI mechanisms was discussed and the direction of future research was indicated. The obtained results could be extended to atomic ions with three valence electrons of $s^2p^1$ configuration.

\begin{acknowledgments}
Support by PL-Grid Infrastructure is acknowledged.
This work was realized under National Science Centre
  (Poland) project Symfonia No. 2016/20/W/ST4/00314 (DKE, AM, JSPB) as well as the OPUS call within the WEAVE programme
2021/43/I/ST3/01142 (JZ). The research has been also supported by a grant from the Priority Research Area (DigiWorld) under the Strategic Programme Excellence Initiative at Jagiellonian University. Views and opinions expressed in this work are, however, those of the authors only and do not necessarily reflect those of the European Union,
nor any other granting authority. Neither the European Union nor any granting authority can be held responsible for them. No part of this work was written with the help of artificial intelligence software.
\end{acknowledgments}

\appendix

\section{\label{App}Symmetry properties of wavefunction for three electrons}

Here we shall discuss the symmetry properties of the spatial part of the wavefunction for three electrons. Let us denote all the quantum numbers needed to describe the state as $Q_k=(q_k,m_{s,k})$, where $m_s$ is a spin projection and $q$ other necessary quantum numbers; by $R_k=(r_k,k)$ let us denote all the variables needed to describe the electron, that is, spatial coordinates and the spin variable $k$. Because electrons are fermions, the wavefunction has to be antisymmetric with respect to exchange of any pair of electrons, and therefore it can be written as a Slater determinant. For three electrons, it reads:
\begin{equation}
    \Psi\propto\left|
    \begin{array}{ccc}
         \Psi_{Q_1}(R_1) & \Psi_{Q_1}(R_2) & \Psi_{Q_1}(R_3)  \\
         \Psi_{Q_2}(R_1) & \Psi_{Q_2}(R_2) & \Psi_{Q_2}(R_3)  \\
         \Psi_{Q_3}(R_1) & \Psi_{Q_3}(R_2) & \Psi_{Q_3}(R_3)
    \end{array}\right|
\end{equation}
Each term of the determinant is composed of three single-electron wavefunctions, each possessing a spatial and spin part, for example:
\begin{multline}
\Psi_{Q_1}(R_1)\Psi_{Q_2}(R_2)\Psi_{Q_3}(R_3) = \\
    \psi_{q_1}(r_1)\phi_{m_{s,1}}(1)
    \psi_{q_2}(r_2)\phi_{m_{s,2}}(2)
    \psi_{q_3}(r_3)\phi_{m_{s,3}}(3) = \\
    \Psi(r_1,r_2,r_3)|m_{s,1}m_{s,2}m_{s,3}\rangle
\end{multline}
In the last line, we collected all the spatial parts and spin parts into two separate functions keeping the order. Other quantum numbers than $m_s$ are omitted in subscript for clarity. When calculating the evolution with our numerical models we need to take only the spatial part with proper symmetry, i.e. when the spin part is symmetric with respect to the exchange of electrons (for example, when it is $|{\rm DDD}\rangle$), the spatial part has to be totally antisymmetric.

In the paper, we discuss the configuration where $m_{s,1}={\rm U}$, $m_{s,2}={\rm U}$ and $m_{s,3}={\rm D}$. So the whole wavefunction has a following form:
\begin{multline}\label{wfunct}
    \Psi\propto \left(\Psi(r_1,r_2,r_3)-\Psi(r_2,r_1,r_3)\right)|{\rm UUD}\rangle \\
        +\left(\Psi(r_2,r_3,r_1)-\Psi(r_3,r_2,r_1)\right)|{\rm DUU}\rangle\\
        +\left(\Psi(r_3,r_1,r_2)-\Psi(r_1,r_3,r_2)\right)|{\rm UDU}\rangle.
\end{multline}
The structure of the wavefunction is simple, it is a sum of three products of spatial and spin functions.

As the Hamiltonian describing the evolution of a system does not have a spin operator part, the spin function remains unchanged, and it is enough to use the spatial function in studying the dynamics. Furthermore, it is not necessary to evolve three separate functions, since proper results are obtained by simple permutation of variables. However, caution should be exercised when discussing the symmetry properties of the partial wavefunctions.

The entire wavefunction is antisymmetric with respect to the exchange of each pair of electrons, but each term on the right of eq.~(\ref{wfunct}) separately behaves differently.
The first term in eq.~(\ref{wfunct}) has spatial part that is antisymmetric with respect to exchange of electrons labeled with 1 and 2 subscripts, and is accompanied with a symmetric spin part with respect to the same exchange. However, if one exchanges electrons 2 and 3 instead, the spatial part is neither symmetric nor antisymmetric, similarly the spin part has no symmetry with respect to that exchange. Likewise the second term in eq.~(\ref{wfunct}) has the property with respect to electrons 2 and 3, and the third term with respect to electrons 1 and 3, respectively.

Now, let us consider exchange of electrons 1 and 3:
in the first term, the spin part changes into the one present in the second term and the spatial part changes accordingly into the one present in the second term but with a minus sign. Here again, if the first term is analyzed only, the spatial part is neither symmetric nor anti-symmetric with respect to that exchange. But the whole wavefunction remains antisymmetric because the discussed exchange of electrons is done in all terms at once, i.e. terms the first and second change into each other with minus sign, the third term just changes sign.

Therefore, for the sake of brevity, let us write the wavefunction in the following form:
\begin{multline}\label{wfunct_fin}
    \Psi\propto \Psi_{12}(r_1,r_2,r_3)|{\rm UUD}\rangle \\
        +\Psi_{23}(r_1,r_2,r_3)|{\rm DUU}\rangle\\
        +\Psi_{13}(r_1,r_2,r_3)|{\rm UDU}\rangle.
\end{multline}
Here, the subscript of each term denotes the pair of coordinates associated with antisymmetry of the spatial part. In our simulations we used only the spatial part of the first term. Let us stress, starting with the spatial function antisymmetric with respect to the change of the electrons 1 and 2 protects the symmetry of the whole wavefunction.

\bibliographystyle{iopart-num}
\bibliography{paper}% Produces the bibliography via BibTeX.

\end{document}